\documentstyle[12pt]{article}
\begin{document}

\textheight 23.0 cm
\textwidth 15.0 cm
\topmargin -1.54 cm
\oddsidemargin 0.8 cm

\def\mach{\vartheta}

\begin{titlepage}

\begin{flushright}
QMW-PH-96-29\\
hep-th/9612208
\end{flushright}

\vspace{3cm}

\begin{center}

{\bf \Large Geometry, Isometries  and Gauging of $ (2,1)$ Heterotic
Sigma-Models}

\vspace{.7cm}

M.\ Abou Zeid and C.\ M.\ Hull

\vspace{.7cm}

{\em Physics Department, Queen Mary and Westfield College, \\
Mile End Road, London E14NS, U.\ K.\ }

\vspace{3cm}

December 1996

\vspace{1cm}

\begin{abstract}

The geometry of (2,1) supersymmetric sigma-models is reviewed and the
conditions under which they have
isometry symmetries are analysed. Certain potentials are constructed that
play an important role in the gauging of such symmetries. The gauged action is
found for a
special class of models.

\end{abstract}

\end{center}

\end{titlepage}

%\begin{document}

\section{(2,1) Geometry}

Heterotic sigma-models with (2,1) supersymmetry have target spaces which are
hermitian manifolds with torsion ~\cite{HW,H1}.   They describe
        the target spaces of heterotic strings with $ (2,1)$   world-sheet
supersymmetry ~\cite{OV}, which   have the remarkable property that different
vacua correspond to the type IIB string and to the membrane of M-theory
{}~\cite{KutMart} and their compactifications, so that they have many potential
applications to the study of   M-theory, string theory and duality.
The construction of the $(2,1)$ string requires that the target space  be four
dimensional with signature  (4,0)
or (2,2)~\cite{OV}, and possess an isometry generated by a null Killing vector,
 which must be gauged~\cite{OV}.   For this reason, it is important to
understand the geometry of gauged (2,1) sigma-models. One approach to the
construction of such gauged models is given in ~\cite{HSP}, but for many
purposes (such as the coupling to supergravity) an approach based on a
conventional superspace formalism is more convenient.
In this note we analyse the geometry associated with  isometry symmetries of
(2,1) sigma-models and construct the potentials that play a central role in the
gauging.
The manifestly supersymmetric gauged action is constructed for a certain class
of isometry
symmetries.
An alternative approach to the gauging of (2,1) models was
discussed in~\cite{HSP}  in which only (1,1) supersymmetry was manifest, but
this
has a number of
disadvantages; for example, the coupling to supergravity is rather inconvenient
in this formalism.
We   instead follow here a more direct route leading to a new form of the
gauged action that is
manifestly (2,1) supersymmetric.
 Many of the results can be applied more generally to (2,p) supersymmetric
sigma-models.
The   gauging of the general (2,1) sigma-models and their applications to
string theory will be addressed in~\cite{AH}.

         The geometric conditions imposed by the requirement of $(2,1)$
world-sheet supersymmetry~\cite{GHR,HW} are (i) that the target manifold $M$ is
a
complex manifold with metric $g_{ij}$ and complex structure  $J^i{}_j$
satisfying
\begin{eqnarray}
J^i{}_j J^j{}_k & = & -\delta^i{}_k \nonumber \\ N^{k}_{ij} & \equiv & J^l{}_i
J^k{}_{[j,l]}
-J^l{}_j J^{k}{}_{[i,l]} = 0
\end{eqnarray}
(ii) that $J^i{}_j$ is covariantly constant
\begin{equation}
\nabla_{i} J^{j}{}_{k} \equiv J^{j}{}_{k,i}+\Gamma^{j}_{il}J^{l}{}_{k}
-\Gamma^{l}_{ik}J^{j}{}_{l} = 0             \label{Jcovconst}
\end{equation}
with respect to the connection
\begin{equation}
\Gamma^{i}_{jk} = \left\{ \begin{array}{c} i \\ jk \end{array} \right\}
+g^{il}H_{jkl}
\end{equation}
which differs from the usual Christoffel connection by the totally
antisymmetric torsion
\begin{equation}
H_{ijk} \equiv \frac{1}{2} \left( b_{ij,k}+b_{jk,i}+b_{ki,j} \right)
\label{defH}
\end{equation}
(iii) the metric $g_{ij}$ is hermitian with respect to the complex structure,
\begin{equation}
g_{ij}J^{i}{}
_{k} J^{j}{}_{l} =g_{kl}  .         \label{ghermJ}
\end{equation}

     In a complex   coordinate system $z^\alpha$,
$\overline{z}^{\overline{\beta}}=(z^\beta )^*$,
  in which the complex structure is
constant and diagonal,
\begin{equation}
J^i{}_{j}  =  i\left( \begin{array}{cc} \delta_{\alpha}^{\beta} & 0  \\
0 & -\delta_{\overline{\alpha}}^{\overline{\beta}} \end{array} \right)  ,
\end{equation}
these conditions imply that the torsion is given by
\begin{equation}
H_{\alpha \beta \overline{\gamma}}=\frac{1}{2}\left( g_{\alpha
\overline{\gamma},
\beta}-g_{\beta \overline{\gamma},\alpha}\right) \ \ , \ \ H_{\alpha \beta
\gamma}=0 ,
\label{Hcomplex}
\end{equation}
while the metric satisfies
\begin{equation}
g_{\alpha [ \overline{\beta},\overline{\gamma}]\delta} -g_{\delta [
\overline{\beta},\overline{\delta}]\alpha} = 0 .
\label{gcomplex}
\end{equation}
The conditions~(\ref{defH}), (\ref{Hcomplex}) and (\ref{gcomplex})
imply the local existence of a vector potential $k_\alpha$ such that
\begin{eqnarray}
g_{\alpha \overline{\beta}} & = & k_{\alpha ,\overline{\beta}}
+\overline{k}_{\overline{\beta} , \alpha } \label{gabbar}\\ b_{\alpha
\overline{\beta}} & = & \overline{k}_{\overline{\beta} ,\alpha} -k_{\alpha
,\overline{\beta}} ,
\label{geom}
\end{eqnarray}
  which defines the geometry locally. If
the torsion $H=0$, the manifold $M$ is K\"{a}hler with $k_\alpha =
\frac{\partial}{\partial z^\alpha }K(z,\overline{z})$ where $K(z,\overline{z})$
is the K\"{a}hler potential and the (2,1) supersymmetric model in fact has
(2,2)
supersymmetry, while for $H \neq 0$ $M$
is a hermitian manifold with torsion.

        The supersymmetric sigma-model can be formulated in (2,1) superspace
in terms of chiral scalar superfields $\varphi^\alpha$  satsfying
\begin{equation}
\overline{D}_{+}\varphi^\alpha =0 \ \ , \ \ D_+
\overline{\varphi}^{\overline{\alpha}}
=0 ,        \label{chiral}
\end{equation}
where $+,-$ are chiral spinor indices and the superspace conventions are as
in~\cite{GHR}. The lowest components of the superfields, $\varphi^\alpha
|_{\theta =0} = z^\alpha $, are the bosonic complex coordinates of the
space-time. The sigma-model action is then given by~\cite{DS}
\begin{equation}
S=i\int d^2 \sigma d \theta_+ d\overline{\theta}_+ d\theta_- \left( k_\alpha
D_-
\varphi^\alpha -\overline{k}_{\overline{\alpha}}D_-
\overline{\varphi}^{\overline{\alpha}} \right) .
\label{21action}
\end{equation}
 The action~(\ref{21action}) is invariant under   the gauge transformation
\begin{equation}
\delta k_\alpha = \rho_\alpha
\label{symm}
\end{equation}
provided $ \rho_\alpha$ satisfies
$ \partial_{\overline{\beta}}\rho_\alpha = i
\partial_\alpha \partial_{\overline{\beta}}\chi  $
for some   arbitrary real  $\chi$. This implies that
$\rho$ is of the form
\begin{equation}
\rho_\alpha = i \partial_\alpha \chi + f_\alpha \ \ , \ \ \bar \partial_
{\bar\beta} f_\alpha=0
\label{ris}
\end{equation}
for some holomorphic $f_\alpha$.  These transformations leave the metric and
torsion invariant, but change $b_{ij}$ by an anti-symmetric tensor gauge
transformation, $\delta b_{ij}=\partial_{[i}\lambda_{j]}$.

\section{ Isometry Symmetries}

        We now consider the isometry symmetries of the target geometry. Let $G$
be a continuous subgroup of the diffeomorphism group of $M$. The action of $G$
on $M$ is generated by vector fields $\xi^i_a$ ($a=1\ldots dim G$) which 
satisfy the Lie bracket algebra
\begin{eqnarray}
[ \xi_a , \xi_b ]^i & \equiv & \xi^j_a \partial_j \xi^i_b - \xi^j_b \partial_j
\xi^{i}_{a} \equiv {\cal L}_a \xi^i_b \nonumber \\ & = & f^{c}_{ab}
\xi^i_c   ,
\label{Liebracket}
\end{eqnarray}
where ${\cal L}_a$ denotes the Lie derivative with respect to $\xi^i_a$ and
$f^{a}_{bc}$ are the structure constants of the group $G$. The
infinitesimal
transformations of the (2,1) sigma model superfields
\begin{equation}
\delta \varphi^i =\lambda^a \xi^i_a        \label{rigidtransf}
\end{equation}
with constant parameters $\lambda^a$ will generate a group of proper
symmetries of the sigma model field equations
if the Lie derivatives with respect to the vector fields $\xi^i_a$ of the
metric
and torsion vanish,
\begin{equation}
({\cal L}_a g )_{ij}=0 \ \ , \ \ ({\cal L}_a H)_{ijk} =0  .
\label{dgdT=0}
\end{equation}
This requires that the $\xi^i_a$ are Killing vectors of the metric $g$,
\begin{equation}
\nabla_{(i}\xi_{j)a}=0  ,    \label{xikill}
\end{equation}
so that
$G$ is a group of isometries of $M$, and that $\xi^i_a H_{ijk}$ is closed,
so that there is a locally defined one-form $u_a$ such that~\cite{HS}
\begin{equation}
\xi^i_a H_{ijk}=\partial_{[j}u_{k]a}   .         \label{dxiT=0}
\end{equation}
For the transformations~(\ref{rigidtransf}) to define a symmetry of the
sigma model action, it is necessary in addition for $u_{ai}$ to be globally
defined.
The one-forms  $u_a$ are only defined up to the addition of an exact piece:
\begin{equation}
u_{ia} \to   u_{ia}+ \partial_i \alpha_a    .  \label{utr}
\end{equation}
Taking the Lie derivative of~(\ref{dxiT=0}), we obtain that
\begin{equation}
D_{iba}\equiv{\cal L}_b u_{ai}-f^{c}_{ba}u_{ic}
\end{equation}
is a closed one-form. If it is exact, it is often
 possible to use the ambiguity~(\ref{utr}) in the
definition of $u_a$ to choose it to be equivariant, i.~e.\ to choose it so that
it transforms as
\begin{equation}
{\cal L}_b u_{ai}=f^{c}_{ba}u_{ic}  .        \label{uequiv}
\end{equation}
However, in general there can be obstructions to choosing an equivariant $u$
which have an interpretation in terms of equivariant cohomology~\cite{HS,OS}.

        While  the conditions~(\ref{dgdT=0}) are sufficient for the isometry
      to be a symmetry of the (1,1) supersymmetric model,
     the isometry will only be compatible with     $ (2,1)$
supersymmetry if the complex structure $J$ is invariant under the
diffeomorphisms generated by the $\xi^i_a$~\cite{H3},
\begin{equation}
({\cal L}_a J)^i{}_j = 0   .        \label{dJ=0}
\end{equation}
Then the $\xi^i_a$ are
Killing vectors which are holomorphic with respect to $J$, so that
\begin{equation}
\partial_{\alpha}\overline{\xi}^{\overline{\beta}}_a = 0  .
\label{xiholo}
\end{equation}
If
the torsion vanishes, then $M$ is K\"{a}hler and for every holomorphic Killing
vector $\xi^i_a$, the
one-form with components $J_{ij}\xi^{j}_{a}$ is closed so that locally there
are
functions $ X_a$ such that
$J_{ij}\xi^{j}_{a}=\partial_i
X_a$; these are  the Killing potentials which play a central role in the
gauging of the
supersymmetric sigma models without torsion~\cite{BW,in&out}.
In complex coordinates, this becomes $\xi_{a \alpha}= -\partial _\alpha X_a$.
When the torsion does
not
vanish, this generalises straightforwardly: if $\xi^i_a$ is a holomorphic
Killing
vector field satisfying~(\ref{dxiT=0}) and~(\ref{dJ=0}), then the one-form with
components $\omega_i \equiv J_{ij}(\xi^{j}_{a}+u_{a}^{j})$ satisfies
 $\partial _{[\alpha}\omega _{\beta]}=0$, so that there are
generalised
complex Killing potentials $Z_a\equiv Y_a+iX_a$ such that
\begin{equation}
\xi_{a \alpha}+u_{a \alpha} =\partial_\alpha Y_a +i\partial_\alpha X_a .
\label{Juxi}
\end{equation}
The $X_a$ and $Y_a$ are locally defined functions and are determined up to
the addition of constants. Note that~(\ref{Juxi})
is invariant under the transformation
  $u_{\alpha a}
\to u_{\alpha a}+\partial_\alpha \alpha_a $ provided that $Y_a$ also transforms
as
$Y_a \to Y_a + \alpha_a$.
It will be useful to absorb $Y$ into $u$, defining
\begin{equation}
u'_{\alpha a} =u_{\alpha a}- \partial_\alpha Y_a  
\label{u'}
\end{equation}
so that $\xi_{a \alpha}+u'_{a \alpha} = i\partial_\alpha X_a $, as in
{}~\cite{HSP}.

        Under the rigid symmetries~(\ref{rigidtransf})
        the variation of the Lagrangian in~(\ref{21action}) is
\begin{equation}
\delta L = i\lambda ^a \left(  {\cal
L}_a
k_\alpha D_{-} \varphi^\alpha  -{\cal
L}_a \overline{k}_{\overline{\alpha}}
D_{-}\overline{\varphi}^{\overline{\alpha}} \right) ,      \label{dS}
\end{equation}
where the Lie derivative of $k_\alpha$ is
\begin{equation}
{\cal L}_a k_\alpha=
\xi^{\beta}_{a}\partial_{\beta}
k_{\alpha}+\xi^{\overline{\beta}}_{a}\partial_{\overline{\beta}}k_{\alpha}
+k_{\beta}\partial_{\alpha}\xi^{\beta}_{a}  .
\label{Liek}
\end{equation}
 In general the
symmetries~(\ref{rigidtransf})
will not leave the action~(\ref{21action}) invariant; they will
leave it invariant only up to  a
 gauge transformation of the form~(\ref{symm}), which requires that
\begin{equation}
{\cal L}_a k_\alpha = i \partial _\alpha \chi_a + \mach _{a \alpha}
\label{kchitheta}
\end{equation}
for some real functions $\chi_a$ and holomorphic one-forms $\mach_{a\alpha}$,
$\partial _{\bar \beta}\mach_{a\alpha}=0$. We will now seek explicit forms for
$\chi,\mach$.

Using the form of the Christoffel symbols
\begin{equation}
\left\{ \begin{array}{c} \gamma \\ \alpha \overline{\beta} \end{array} \right\}
= -g^{\gamma \overline{\delta}}H_{\alpha \overline{\beta}\overline{\delta}} \ \
,
\ \ \left\{ \begin{array}{c} \overline{\gamma} \\ \alpha \overline{\beta}
\end{array} \right\} = g^{\overline{\gamma}\delta}H_{\delta \alpha
\overline{\beta}} \ \ ,
\end{equation}
we find that the Killing equation~(\ref{xikill}) becomes
\begin{equation}
0 = \nabla_{(\alpha}\overline{\xi}_{\overline{\beta} )a}=\partial_{(\alpha}
\overline{\xi}_{\overline{\beta} )a}-H_{\alpha \overline{\beta}}^{\, \, \,
\overline{\gamma}}
\overline{\xi}_{\overline{\gamma}a} +H_{\alpha \overline{\beta}}^{\, \, \, \gamma}
\xi_\gamma  .
\end{equation}
Comparing with~(\ref{dxiT=0}), which yields
\begin{equation}
\partial_{[\alpha}\overline{u}_{\overline{\beta}]a} = H_{\alpha 
\overline{\beta}}^{\, \,
\, \, \, \, \gamma}\overline{\xi}_{\gamma a} + H_{\alpha 
\overline{\beta}}^{\, \, \, \, \,
\,
\overline{\gamma}} \xi_{\overline{\gamma} a}  ,
\end{equation}
we find the relation
\begin{equation}
2H_{\alpha \overline{\beta}}^{\, \, \, \, \, \, \gamma}\xi_{\gamma a} =
\partial_{[\alpha}
\overline{u}_{\overline{\beta}]a}-\partial_{(\alpha}
\overline{\xi}_{\overline{\beta})a}  .
\label{2H1}
\end{equation}
Furthermore, eq.~(\ref{Juxi}) gives
\begin{equation}
\xi_{\alpha a }= \partial_{\alpha} (Y_a +iX_a ) -u_{\alpha a} ,
\label{xiYX}
\end{equation}
which implies that~(\ref{2H1}) can be rewritten in the following two equivalent
ways
\begin{eqnarray}
2H_{\alpha \overline{\beta}}^{\, \, \, \, \, \, \gamma}\xi_{\gamma a} & = &
\partial_\alpha
(\overline{u}_{\overline{\beta}a}-\partial_{\overline{\beta}}Y_a ) \nonumber \\ & = &
-\partial_\alpha (\overline{\xi}_{\overline{\beta}a}+i\partial_{\overline{\beta}}X_a )  .
\label{2H23}
\end{eqnarray}
Moreover, it   follows from~(\ref{xiholo}) and eqs.~(\ref{Hcomplex})
and~(\ref{gabbar}) that
\begin{equation}
H_{\alpha \overline{\beta}}^{\, \, \, \, \, \, \gamma}\xi_{\gamma a} =
\partial_\alpha
\left( \overline{\xi}^{\overline{\gamma}}_a 
\overline{k}_{[\overline{\beta},\overline{\gamma}]}
\right)  .
\label{Hd()}
\end{equation}
Hence, substituting this in~(\ref{2H23})  and integrating,  we find
\begin{equation}
-\left( \overline{\xi}_{\overline{\beta}a}+i\partial_{\overline{\beta}}X_a \right)
=2\overline{\xi}^{\overline{\gamma}}_{a}
\overline{k}_{[\overline{\beta},\overline{\gamma}]}
-\overline{\mach}_{\overline{\beta}a}      \label{intl1}
\end{equation}
for some antiholomorphic function $\overline{\mach}_{\overline{\alpha}a}$. The
holomorphy of
\begin{equation}
\mach_{\alpha a}= 2\xi^{\gamma}_a \partial_{[\gamma}k_{\alpha ]} +\xi_{\alpha
a}
-i\partial_\alpha X_a  ,
\label{theta}
\end{equation}
which follows from the above construction, can also be checked by
direct calculation using eqs.~(\ref{xiholo}) and~(\ref{xiYX}).

        Similarly,~(\ref{2H23}) also yields an expression for
$u_{\alpha a}$: subtituting~(\ref{xiYX}) in~(\ref{theta}), we
find
\begin{equation}
u_{\alpha a} = \partial_\alpha Y_a +2\xi^{\gamma}_a \partial_{[\gamma}
k_{\alpha ]}- \mach_{\alpha a} .
\label{u}
\end{equation}
It is easily checked that the expression~(\ref{u}) of $u_{\alpha a}$ is
compatible with the geometric condition~(\ref{dxiT=0}).

        Note that the right hand side of~(\ref{xiYX}) is invariant  under
$u_{\alpha a}
\to u_{\alpha a}+\partial_\alpha \alpha_a $  and $Y_a \to Y_a + \alpha_a$, as
it should be.
Absorbing $Y$ into $u$, as in~(\ref{u'})
   the one-form
\begin{equation}
D'_{iba} \equiv {\cal L}_i u'_{i a} -f^{c}_{ba}u'_{ic}
\label{defD}
\end{equation}
is closed, which implies the local existence of a real potential $E_{ba}$ such
that
\begin{equation}
D'_{ba\alpha} =i\partial_\alpha E_{ba}  .
\label{D&E}
\end{equation}
(note that $E_{ba}$ is only defined up to the addition of real constants).
In turn, the
potential $E_{ba}$ is determined by the imaginary part of the
generalised Killing potential. This is seen by taking the Lie derivative of
eq.~(\ref{xiYX}) and integrating, which yields
\begin{equation}
E_{ba} = {\cal L}_b X_a -f^{c}_{ba}X_c +e_{ba}
\label{E&X}
\end{equation}
where the $e_{ba}$ are real constants which we henceforth absorb  into the
definition
of $E_{ba}$.

        Note that the ambiguity $X_a \to X_a + C_a$ in the definition of
$X_a$ (for some constant   $C_a$) does not affect
$\mach$. Under the
transformations~(\ref{symm}), (\ref{ris}), both $\mach$
and $\chi$ undergo certain shifts, as can be
checked using the forms~(\ref{theta}) and~(\ref{defchi}).
        Now, using the relations~(\ref{geom}), (\ref{xiholo})
and~(\ref{theta}),
we find that the Lie derivative of $k_\alpha$ can be written in the following
way:
\begin{eqnarray}
 {\cal L}_a k_\alpha  & = &
\overline{\xi}^{\overline{\beta}}_a \left( \partial_{\overline{\beta}} k_\alpha +
\partial_\alpha \overline{k}_{\overline{\beta}} \right) -\partial_\alpha \left(
\overline{\xi}^{\overline{\beta}}_a 
\overline{k}_{\overline{\beta}}-\xi^{\beta}_a k_\beta \right)
+2\xi^{\beta}_a \partial_{[\beta}k_{\alpha ]} \nonumber \\ & = &
i\partial_\alpha
\chi_a +\mach_{\alpha a}
\label{rho}
\end{eqnarray}
and we have found that $\mach $ is given by~(\ref{theta}), while
\begin{equation}
\chi_a \equiv X_a + i\left( \overline{\xi}^{\overline{\beta}}_a 
\overline{k}_{\overline{\beta}}
-\xi^{\beta}_a k_\beta \right)  .
\label{defchi}
\end{equation}

        Further information into the relation of the isometry subgroup $G$ of
$M$
to its geometry can be obtained by deriving the action of the Lie bracket
algebra on $k_\alpha$. First, note that~(\ref{rho}) and~(\ref{defchi}) imply
\begin{eqnarray}
{\cal L}_b \left( \chi_a -X_a \right) & = & f^{c}_{ba}\left( \chi_c -X_c
\right)
+ i\left[ \overline{\xi}^{\overline{\beta}}_a 
\left( -i\partial_{\overline{\beta}}\chi_b +
\overline{\mach}_{\overline{\beta}b} \right) -\xi^{\beta}_a \left(
i\partial_\beta \chi_b +\mach_{\beta b} \right) \right] \nonumber \\
& = & f^{c}_{ba}\left( \chi_c -X_c \right) +{\cal L}_a \chi_b +i\left(
\overline{\xi}^{\overline{\beta}}_a 
\overline{\mach}_{\overline{\beta}b}-\xi^{\beta}_a
\mach_{\beta b}\right)
\end{eqnarray}
so that
\begin{equation}
{\cal L}_b \chi_a -{\cal L}_a \chi_b -f^{c}_{ba}\chi_c = {\cal L}_b X_a
-f^{c}_{ba}X_c +i\left( \overline{\xi}^{\overline{\beta}}_a 
\overline{\mach}_{\overline{\beta}b}
-\xi^{\beta}_a \mach_{\beta b} \right)  .
\label{chialg}
\end{equation}
Then, taking the Lie derivative of~(\ref{rho}) with respect to the isometry
generated by $\xi^{\alpha}_b$ and substracting the resulting equation with
group
indices interchanged, we find
\begin{eqnarray}
[ {\cal L}_b ,{\cal L}_a ] k_\alpha & = & f^{c}_{ba} {\cal L}_c k_\alpha
+i\partial_\alpha \left( {\cal L}_b \chi_a -{\cal L}_a \chi_b -f^{c}_{ba}\chi_c
\right) \nonumber \\
& & +\left( {\cal L}_b \mach_{\alpha a} -{\cal L}_a \mach_{\alpha b}
-f^{c}_{ba}\mach_{\alpha c} \right) ,
\end{eqnarray}
which can be rewritten as
\begin{eqnarray}
[ {\cal L}_b ,{\cal L}_a ] k_\alpha & = & f^{c}_{ba} {\cal L}_c k_\alpha
+i\partial_\alpha \left( {\cal L}_b X_a -f^{c}_{ba}X_c \right) \nonumber \\
& & +\partial_\alpha \left( \xi^{\beta}_{a} \mach_{\beta  b} \right) +\left(
{\cal L}_b \mach_{\alpha a} -{\cal L}_a \mach_{\alpha b}
-f^{c}_{ba}\mach_{\alpha c} \right)
\label{Liealg-}
\end{eqnarray}
upon substituting eq.~(\ref{chialg}). On the other hand, the Lie derivatives
  satisfy the Lie algebra of $G$, so that
\begin{equation}
[ {\cal L}_a ,{\cal L}_b ] k_\alpha = f^{c}_{ab} {\cal L}_c k_\alpha
\label{Liealg}
\end{equation}
Thus the sum of the last three terms on the right hand side
of~(\ref{Liealg-}) must vanish. We will now show that this is indeed the case.

        First, taking the Lie derivative of $\mach_{\alpha a}$
in~(\ref{theta}) with respect to $\xi^i_b$, we find
\begin{equation}
{\cal L}_b \mach_{\alpha a} = f^{c}_{ba}\mach_{\alpha c} +2\xi^{\gamma}_a
\partial_{[\gamma}\mach_{\alpha ]b} -D'_{ba} ,
\label{Lietheta}
\end{equation}
where we have used eq.~(\ref{rho}), the relation~(\ref{E&X}) and the
definition~(\ref{D&E}). The second
term on the right hand side of~(\ref{Lietheta}) can be rewritten as
\begin{eqnarray}
2\xi^{\gamma}_a \partial_{[\gamma}\mach_{\alpha ]b} & = & \xi^{\gamma}_a
\partial_\gamma \mach_{\alpha b} +\mach_{\gamma b}\partial_\alpha \xi^{\gamma
a}
-\partial_\alpha \left( \xi^{\gamma}_a \mach_{\gamma b}\right) \nonumber \\ &
= &
{\cal L}_a \mach_{\alpha b} -\partial_\alpha \left( \xi^{\gamma}_a
\mach_{\gamma
b}\right)
\label{xidtheta}
\end{eqnarray}
using the holomorphy of $\mach_{\alpha b}$. Substituting~(\ref{xidtheta})
in~(\ref{Lietheta}), we find the relation
\begin{equation}
{\cal L}_b \mach_{\alpha a}-{\cal L}_a \mach_{\alpha b} = f^{c}_{ba}
\mach_{\alpha c} -\partial_\alpha \left( \xi^\gamma_a \mach_{\gamma b}
+iE_{ba} \right)  .
\label{thetequiv-}
\end{equation}
Inserting~(\ref{E&X}) and~(\ref{thetequiv-}) in~(\ref{Liealg-}), we find
that the sum of the last three terms on the right hand side of~(\ref{Liealg-})
explicitly cancels when~(\ref{D&E}) and~(\ref{E&X}) are used, so
that~(\ref{Liealg-}) indeed reduces to~(\ref{Liealg}).

        Another important consequence of~(\ref{thetequiv-}) follows from
symmetrization with respect to group indices: this yields
\begin{equation}
\partial_\alpha \left( \xi^{\gamma}_{(a}\mach_{\gamma b)} +iE_{(ba)}
\right) = 0 ,
\label{ddab}
\end{equation}
which upon integration implies that
\begin{equation}
\hat{d}_{(ab)} \equiv -\xi^{\gamma}_{(a}\mach_{\gamma b)} +iE_{(ba)}
\label{hatd}
\end{equation}
is an antiholomorphic function, $\hat{d}_{ab}=\hat{d}_{ab}(\overline{z})$.
Then, defining $c_{(ab)}$ as the real part of $\hat{d}_{(ab)}$, we find
\begin{equation}
c_{(ab)} \equiv \hat{d}_{(ab)}+\overline{\hat{d}}_{(ab)} =- \xi^{i}_{(a}
\mach_{ib)}
{}.
\label{defd}
\end{equation}
We now show that the $c_{(ab)}$ are real constants. Substituting the
explicit expression~(\ref{theta})
for $\mach_{\alpha a}$ and using the relation~(\ref{u}), we find
\begin{equation}
\xi^{\gamma}_a \mach_{\gamma b}   =   2\xi^{\gamma}_a \xi^{\beta}_b
\partial_{[
\beta}k_{\gamma ]} - \xi^{\gamma}_a  u'_{\gamma b} .
\end{equation}
Then, symmetrization with respect to group indices yields
\begin{equation}
\xi^{\gamma}_{(a}\mach_{\gamma b)} = -\xi^{\gamma}_{(a}u'_{|
\gamma | b)}  ,
\end{equation}
Hence, we find that~(\ref{defd}) can be rewritten as
\begin{equation}
c_{(ab)} = \xi^{i}_{(a}u'_{ib)} ,
\end{equation}
This is precisely the definition of $c_{(ab)}$ given in
ref.~\cite{HS,HSP}, where it was shown that they are
  real constants  whose vanishing is   a
necessary condition for the gauging of the sigma model to be
possible~\cite{HS,HSP}.

        The equivariance condition on the imaginary part of the generalised
Killing potential,
\begin{equation}
{\cal L}_b X_a = f^{c}_{ba}  X_c  ,
\label{Xequiv}
\end{equation}
was found in~\cite{HS,HSP} to be another necessary condition for the gauging of
the
isometries generated by the $\xi^i_a$ to be possible. If~(\ref{Xequiv}) holds ,
then it
follows from~(\ref{E&X}) that the potential $E_{ba}$ defined in~(\ref{D&E})
is a constant and can be chosen to vanish,
\begin{equation}
E_{ba} = 0  ,        \label{Eis0}
\end{equation}
and that eqs.~(\ref{thetequiv-}), (\ref{ddab}) and~(\ref{hatd}) simplify.
The equations~(\ref{defD}), (\ref{D&E}) then imply that
   $u'$ is equivariant.

        Summarizing, the action of a group  $G$ generated by the vector fields
$\xi^i_a$ as in~(\ref{rigidtransf}) is a symmetry provided the $\xi^i_a$ are
holomorphic
Killing vectors, i.~e.\ eqs.~(\ref{xiholo}) and~(\ref{xikill}) hold, so that
the
metric and complex structure are   invariant, and in addition the
torsion is invariant, i.~e.\
eqs.~(\ref{dgdT=0}) and~(\ref{dJ=0}) hold. In general, the isometry symmetries
will not leave the action~(\ref{21action}) invariant, but will leave it
invariant up to a gauge transformation of the form~(\ref{symm}). The geometry
and Killing potentials then determine the quantity ${\cal L}_ak_{\alpha  }$
appearing in the gauge transformation to take the form~(\ref{rho}), with $\chi
,
\mach$ as in~(\ref{defchi}) and~(\ref{theta}). It is then found that the
Lie bracket algebra of the function $k_\alpha$ which determines the geometry
closes. Also, the quantities defined in~(\ref{defd}) are the
real constants $c_{(ab)}$ of
ref.~\cite{HSP}.
When the imaginary part of the generalised Killing potential is chosen to be
equivariant, i.~e.\
when~(\ref{Xequiv}) holds, it is found that the potential $E_{ba}$ defined
in~(\ref{D&E}) vanishes. Then the one-forms $u'_a$ defined in~(\ref{u'})
are equivariant and the geometry simplifies. We note the result of
ref.~\cite{HSP}, where it was shown that the equivariance
condition~(\ref{Xequiv})
on the imaginary part of the generalised Killing potential must hold
in order for the gauging of the supersymmetric sigma model to be possible.

        The discussion given here also applies to the geometry and isometries
of the target space of (2,0) heterotic strings. The corresponding formulae can
be obtained from those given in the foregoing by appropriate truncation of the
(2,1) superfields.

\section{Gauging the Isometries}

We now turn to the gauging of the (2,1) sigma-model. The aim is
to promote the rigid isometry symmetries~(\ref{rigidtransf}) to local ones in
which the scalar fields
transform as
\begin{equation}
\delta \varphi^\alpha = \Lambda^a \xi^{\alpha}_a \ \ , \ \ \delta
\overline{\varphi}^{\overline{\alpha}} = \overline{\Lambda}^a
\overline{\xi}^{\overline{\alpha}}_{a} ,
\label{iso}
\end{equation}
where the parameters
 $\Lambda^a,\bar \Lambda^a$ satisfy the  chirality conditions
\begin{equation}
\overline{D}_{+}\Lambda^a = 0 \ \ , \ \ D_+ \overline{\Lambda}^a = 0 .
\end{equation}
Under a finite transformation,
\begin{equation}
\varphi \rightarrow \varphi ' = e^{L_{\Lambda \cdot \xi}}\varphi \ \ , \ \
\overline{\varphi} \rightarrow \overline{\varphi} '
= e^{L_{\overline{\Lambda} \cdot \overline{\xi}}}\overline{\varphi}  ,
\label{isoloc}
\end{equation}
where
\begin{equation}
\Lambda \cdot \xi \equiv \Lambda^a \xi^{\alpha}_{a}\frac{\partial}{\partial
\varphi^{\alpha}}
\end{equation}
and $L_{\Lambda \cdot \xi}\varphi^{\alpha}$ denotes the action of the
infinitesimal diffeomorphism with parameter $\Lambda \cdot \xi$,
\begin{equation}
L_{\Lambda \cdot \xi}\varphi^{\alpha} \equiv   \Lambda \cdot \xi^{\alpha} ,
\end{equation}
and acts on tensors as the Lie derivative with respect to $\Lambda \cdot \xi$.
To construct the gauged action, we couple the sigma-model to the (2,1)
supersymmetric gauge multiplet~\cite{HSP}. The constraints for this multiplet
can be solved to express the superconnections in terms of a prepotential $V$,
which transforms as
\begin{equation}
e^V \rightarrow e^{V '}=e^\Lambda e^V e^{-\overline{\Lambda}} ,
\label{transfV}
\end{equation}
and a spinorial connection $A_-^a$, with the infinitesimal  transformation
\begin{equation}
\delta A_{-}^{a}  =   D_- \Lambda^{a} +\left[ A_- ,\Lambda \right]^a  .
\label{transfAs}
\end{equation}
We choose a chiral representation in which the right-handed covariant
derivatives are
\begin{equation}
\overline{\nabla}_{+} =\overline{D}_{+} \ \ , \ \ \nabla_+ = e^V D_+ e^{-V} ,
\label{chirrep}
\end{equation}
while the left-handed covariant derivative is defined by
\begin{equation}
\nabla_- \varphi^\alpha = D_- \varphi^\alpha -A_{-}^a \xi^{\alpha}_{a} .
\label{nabla-}
\end{equation}

     Now let us define (following ~\cite{in&out})
\begin{equation}
\tilde{\varphi} = e^{L_{V\cdot \overline{\xi}}}\overline{\varphi} ,
\label{deffitilde}
\end{equation}
where
\begin{equation}
L_{V\cdot \overline{\xi}} = V^a \overline{\xi}^{\overline{\alpha}}_a
\frac{\partial}{\partial \overline{\varphi}^{\overline{\alpha}} }  .
\label{LV}
\end{equation}
Then the   fields $\varphi ,\tilde{\varphi}$ satisfy the covariant
chiral constraints
\begin{equation}
\overline{\nabla}_+ \varphi^\alpha = 0 \ \ , \ \ \nabla_+
\tilde{\varphi}^{\overline{\alpha}} = 0  ,
\end{equation}
and transform as
\begin{equation}
\delta \varphi^\alpha = \Lambda^a \xi^\alpha_a \ \ , \ \ \delta
\tilde{\varphi}^{\overline{\alpha}}_a =\Lambda^a
\tilde{\xi}^{\overline{\alpha}} (\tilde{\varphi})  .
\label{dfifitilde}
\end{equation}
Note that the $\tilde \varphi $ transformation involved the parameter $\Lambda$
while that for $\bar \varphi$ involved $\bar \Lambda$.
The left-handed covariant derivative of $\tilde \varphi$ is
\begin{equation}
\nabla_- \tilde\varphi^{\bar \alpha }= D_- \tilde \varphi^{\bar \alpha }
-A_{-}^a \tilde{\xi}^{{\bar \alpha }}_{a}(\tilde \varphi)  .
\label{nablatild-}
\end{equation}

        In general, the potential $k_\alpha$ is only gauge invariant up to a
transformation of the form~(\ref{symm}), so that its Lie derivative is given
by~(\ref{kchitheta}) for some $\chi, \mach$.
Consider first the special case in which ${\cal L}_a k_\alpha=0$, so that the
action~(\ref{21action}) is invariant under the rigid
transformations~(\ref{rigidtransf}). Then the gauged sigma-model is obtained by
minimal
coupling. This coupling is achieved by replacing $\bar \varphi$
with $  \tilde \varphi$ and replacing the supercovariant derivative $D_-$
with the gauge covariant derivative $\nabla_-$ defined in~(\ref{nabla-})
and~(\ref{nablatild-}). This gives the Lagrangian
\begin{equation}
L_0 = i\left( k_\alpha (\varphi ,\tilde{\varphi})\nabla_- \varphi^\alpha
-\tilde{k}_{\overline{\alpha}}(\varphi ,\tilde{\varphi}) \nabla_-
\tilde{\varphi}^{\overline{\alpha}} \right)  . \label{L0}
\end{equation}
This is indeed invariant under the transformations (\ref{transfV}),
(\ref{transfAs}) and~(\ref{dfifitilde}) provided  ${\cal L}_a k_\alpha=0$.

        Consider now the more general case in which~(\ref{kchitheta}) holds,
but with $\mach=0$, so that
\begin{equation}
{\cal L}_a k_\alpha = i \partial _\alpha \chi_a .
\end{equation}
The action based on~(\ref{L0}) is no longer gauge invariant;
using~(\ref{transfV}), (\ref{transfAs}), (\ref{nabla-}) and the
infinitesimal variation of the fields~(\ref{dfifitilde}), we find
\begin{equation}
\delta L_0 =i\Lambda^a D_- \chi_a (\varphi ,\tilde{\varphi})
+\Lambda^a A_{-}^{b} \left( \partial_\alpha \chi_a \xi^{\alpha}_{b}
+\partial_{\overline{\alpha}} \chi_a A_{-}^{b}
\tilde{\xi}^{\overline{\alpha}}_b \right) (\varphi ,\tilde{\varphi}) .
\label{dL0}
\end{equation}
This can be cancelled by  adding the following term to $L_0$:
\begin{equation}
\hat{L}_0 =-A_{-}^{a}  \chi_{a}   (\varphi ,\tilde{\varphi})  .
\label{L0hat}
\end{equation}
The definition~(\ref{defchi}) implies that the terms multiplying the gauge
connection $A_-$ combine to yield the generalised Killing potential $X$:
\begin{eqnarray}
L_{g}^{(0)} & = & L_0 +\hat{L}_0 \nonumber \\ & = & i\left( k_\alpha D_-
\varphi^\alpha -\tilde{k}_{\overline{\alpha}}D_-
\tilde{\varphi}^{\overline{\alpha}} \right) (\varphi , \tilde{\varphi})
-A_{-}^{a}  X_{a}   (\varphi ,\tilde{\varphi}) .
\label{Lgauge0}
\end{eqnarray}
The Lagrangian $L_{g}^{(0)}$ in~(\ref{Lgauge0}) is the full
gauge-invariant action for the gauged (2,1) model in the special case where
$\mach =0$ provided that the generalised Killing potential $X$ transforms
covariantly under the isometries~(\ref{iso}), i.~e.\
\begin{equation}
\delta X_a =f^{c}_{ab}\Lambda^b X_c  .
\label{XeqLambda}
\end{equation}
To see this, note that the variation of the first term in~(\ref{Lgauge0}) is
given by
\begin{eqnarray}
\delta \left[ i\left( k_\alpha D_- \varphi^\alpha
-\tilde{k}_{\overline{\alpha}}
D_- \tilde{\varphi}^{\overline{\alpha}}\right) (\varphi ,\tilde{\varphi})
\right]
& = & -\Lambda^a D_- \chi_a (\varphi ,\tilde{\varphi}) -iD_- \Lambda^a
\left( \tilde{\xi}^{\overline{\alpha}}_a k_{\overline{\alpha}}-\xi^{\alpha}_a
k_\alpha \right) (\varphi ,\tilde{\varphi}) \nonumber \\ & = & -D_- \Lambda^a
X_a (\varphi ,\tilde{\varphi}) ,
\label{dL0D}
\end{eqnarray}
the manipulations being similar to those which lead to the
expression~(\ref{dL0}); the last identity follows upon integrating by parts,
discarding a surface term and using the definition~(\ref{defchi}) (notice that
the
second term on the right-hand side of~(\ref{defchi}) has cancelled). On the
other hand, the variation of the second term in~(\ref{Lgauge0}) is
\begin{equation}
\delta \left[ - A_{-}^a X_a  (\varphi ,\tilde{\varphi})
\right] = -D_- \Lambda^a X_a (\varphi ,\tilde{\varphi})
-A_{-}^a \left( \delta X_a +f^{c}_{ba}\Lambda^b X_c \right) (\varphi ,
\tilde{\varphi}) ,
\label{dL0hat}
\end{equation}
where we have used the variations~(\ref{transfAs}) of the gauge connection.
Adding
(\ref{dL0D}) and~(\ref{dL0hat}), a cancellation occurs, and one is left with
\begin{equation}
\delta L_{g}^{(0)} = -A_{-}^{a} \left( \delta X_a + f^{c}_{ba}\Lambda^b X_c
\right)
(\varphi ,\tilde{\varphi}) ,
\label{dLg0}
\end{equation}
which vanishes if the equivariance condition~(\ref{XeqLambda}) is satisfied.
Furthermore, it is easily seen that~(\ref{XeqLambda}) must be imposed if the
variation~(\ref{dLg0}) is to vanish, as no simple Lorentz-invariant object
can be constructed with variation given by the gauge connection.
As seen in the last section, the condition~(\ref{XeqLambda}) implies the
equivariance of $u'$ while $c_{(ab)}=0$ as a result of ~(\ref{defd})
 and the assumption that $\mach=0$. Thus the obstructions to gauging found
in~\cite{HS,HSP} are all absent.

        Summarizing, we find that, in the special case where $\mach =0$, the
action~(\ref{21action}) for the (2,1) model can be gauged provided the same
geometric
condition as that found in ref.~\cite{HSP} is satisfied, namely the
equivariance
of the generalized Killing potential $X$. Moreover, if~(\ref{XeqLambda}) holds,
then the gauged (2,1) sigma-model action in this case is the superspace
integral
of the gauge invariant Lagrangian~(\ref{Lgauge0}). The general case in which
$\mach \ne 0$ and
the Lie derivative of $k_\alpha$ is given by~(\ref{kchitheta}) is more
complicated and will be treated in~\cite{AH}, using the methods
of~\cite{in&out}.

\vspace{1.5cm}

{\bf Acknowledgements}
\\
\\
        The work of M.\ A.\ is supported in part by the Board of the Swiss
Federal Institutes of Technology, by the Fonds National Suisse and by the
Overseas Research Scheme.

\end{document}